\documentclass[prb,twocolumn]{revtex4-1}
\usepackage{graphicx}
\usepackage{hyperref}
\usepackage{amssymb}
\usepackage{dcolumn}
\usepackage{float}
\usepackage{bm}
\usepackage{multirow}
\usepackage{booktabs}
\usepackage[T1]{fontenc}
\usepackage[utf8]{inputenc}
\usepackage{lmodern}
\usepackage{color}
\usepackage{makecell}
\usepackage{tikz}
\usepackage{pgfplots}
\usepackage{mathtools}
\usepackage{amsmath}

\newcolumntype{M}[1]{>{\centering\arraybackslash}m{#1}}

\newcommand{\bs}{\boldsymbol}
\DeclareMathAlphabet{\bi}{OML}{cmm}{b}{it}

\def\be{\begin{equation}}
\def\ee{\end{equation}}
\def\bearr{\begin{eqnarray}}
\def\eearr{\end{eqnarray}}

\begin{document}
\title{Direction dependent giant optical conductivity in 2D \textit{semi}-Dirac materials}
\bigskip
\author{Alestin Mawrie and Bhaskaran Muralidharan}
\normalsize
\affiliation{Department of Electrical Engineering, Indian Institute of Technology Bombay, Powai, Mumbai-400076, INDIA}
\begin{abstract}
We show that the gap parameter in \textit{semi}-Dirac material induces a large degree of sensitivity for inter-band optical conductivity with respect to the polarization direction. The optical conductivity reveals an abruptly large value at a certain frequency for light along a particular polarization direction while it is significantly suppressed along the direction orthogonal to the former. The direction-dependent optical conductivity may, in turn, be used to uniquely predict the dispersive nature of the 2D \textit{semi}-Dirac materials, in addition to other possible applications in the field of transparent conductors.
\end{abstract}
\pacs{78.67.-n, 72.20.-i, 71.70.Ej}
\maketitle
Low energy excitations with massless Dirac particle behavior are characteristics of some illustrious 2D and quasi-2D materials such as graphene\cite{graphene1,graphene2}, silicene\cite{silicene1}, MoS$_2$ \cite{MoS21,MoS22,MoS23}, $8$-$Pmmn$ borophene \cite{boropheneElec}, to name a few. 
The massless Dirac particles have a low energy band dispersion, which is linear in all {\bf k}-space directions (also known as Dirac cone), with particle and hole states lying respectively above and below a nodal point called the Dirac node. 
The Dirac-cone in the dispersion spectrum controls the various low energy properties such as specific heat\cite{specific}, suppression of back-scattering\cite{backsc,borophene-op}, transport properties such as optical conductivity\cite{sarma} and magnetic field responses of such 2D materials. 
\\
\indent
Recently, a distinct class of 2D Dirac materials called \textit{semi}-Dirac (SD) materials has been discovered in materials or systems such as TiO$_2$/V$_2$O$_3$ nanostructure\cite{ms1}, 
dielectric photonic systems\cite{ms4} and hexagonal lattices in presence of a magnetic field\cite{ms3}. SD material has a unique low energy dispersion, which is quadratic in a given direction and linear in the orthogonal direction with respect to the former. The band anisotropy in SD materials was found to be stable against weak short-range interaction while there is a direct Dirac-semimetal to band insulator transition for stronger interaction\citep{BRoy}.
The low energy Hamiltonian that features the anisotropic band dispersion in SD materials goes as\cite{Hamilt1,Hamilt2}, 
\begin{eqnarray}\label{Hamil}
H_0=\textbf{g}(\textbf{k})\cdot\boldsymbol{\sigma},
\end{eqnarray}
where ${\bf g}({\bf k})\equiv (\alpha k_x^2-\delta_0,vk_y)$ with $\alpha$, $\delta_0$ and $v$ being the inverse of quasiparticle mass along $x$-direction, the system gap parameter and Dirac quaisparticle velocity along $y$-direction, respectively and $\boldsymbol{\sigma}\equiv(\sigma_x,\sigma_y)$ are the $2\times 2$ Pauli's matrices. A term Type-$I$ SD for the above Hamiltonian was coined by Huang, \textit{et. al.}\cite{Huang}, which differs from another type-$II$ SD Hamiltonian since the latter also described the emergence of Chern insulating states in the super-crystal (TiO$_2$)$_5$/(VO$_2$)$_3$\cite{Huang}. In this paper, we refer only to the type-$I$ SD Hamiltonian.
\begin{figure*}[!htbp]
	\begin{center}\leavevmode
		\includegraphics[width=180mm,height=47.9mm]{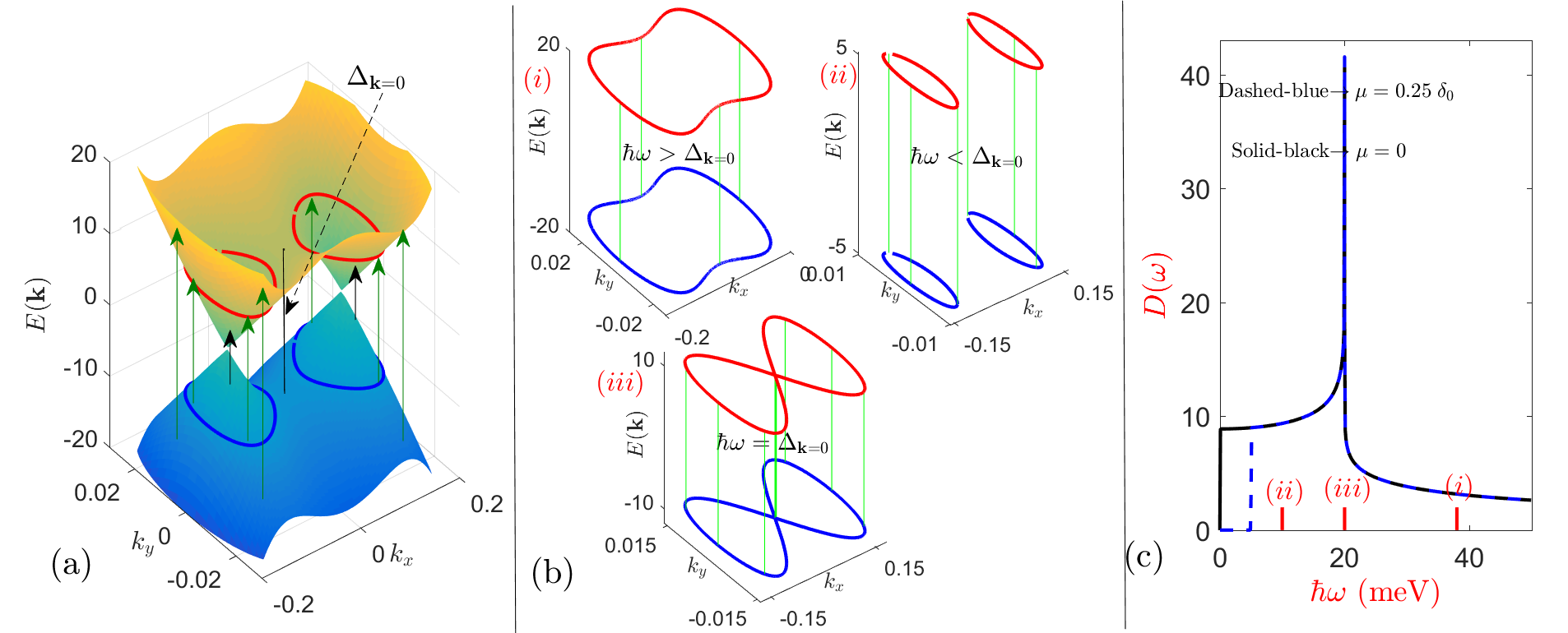}
		\setlength{\belowcaptionskip}{-12pt}
		\caption{($a$). Plot of the band dispersion for SD materials. The Fermi contour is shown by red curves. The arrows show the transitions across the particle and hole state with allowed transitions (green color) and disallowed transitions (black color). The length of the arrow indicates the light frequency. ($b$) Plots depicting the different available inter-band transitions across the three different constant energy contours which indicate the three possibilities of the magnitute of $\hbar\omega$. The van-Hove singularity occurring when $\hbar\omega=\Delta_{{\bf k}=0}$ which leads to maximum JDOS at this frequency as shown in $(c)$ and indicated by the point ($iii$).}
		\label{Fig1}
	\end{center}
\end{figure*}
The eigensystem of $H_0$ are $
\varepsilon_{\lambda}({\bf k}) = \lambda\sqrt{g_y^2+g_x^2}$
and
$
\psi_{\bf k}^{\lambda}({\bf r}) = \frac{e^{i\textbf{k} \cdot \textbf{r}}}{\sqrt{2}}
\begin{pmatrix}
1,&
\lambda i\frac{-ig_x+g_y}{\sqrt{g_y^2+g_x^2}}
\end{pmatrix}^\mathcal{T},
$ respectively,
where $\lambda = +/-$ denotes the conduction/valence band and $\mathcal{T}$ the transpose. The energy separation between the valence and conduction bands is $\Delta_{\bf k}=\varepsilon_+({\bf k})-\varepsilon_-({\bf k})=2\sqrt{g_y^2+g_x^2}$. 
\\
\indent 
The gap parameter in the Hamiltonian given in Eq. (\ref{Hamil}), can be $(i)$ $\delta_0=0$, which represents gapless spectrum
; $(ii)$ $\delta_0<0$, which gives a trivial insulating phase with a non-zero energy gap; and $(iii)$ $\delta_0>0$, which gives the 2D SD gapless states that uniquely possess two nodal points exactly at $(\pm k_0,0)$, with $k_0=\sqrt{{\delta_0}/{\alpha}}$ (Fig. [\ref{Fig1}]). The gapless states with $\delta_0>0$, are stable against short-range fermion-fermion interaction or impurities\cite{JPCM30}.
The electron and hole states are degenerated at the two nodal points and separated by a gap $\Delta_{\bf k}$ elsewhere. 
There are also theoretical predictions of photoinduced topological phase transition and gap opening at the two nodal points at high momentum of the radiation \cite{kush,Firoz_SD}. The plot of band dispersion with respect to Eq. (\ref{Hamil}), 
with the gap parameter set to $\delta_0>0$ is shown in Fig. [\ref{Fig1}($a$)].
\\
\indent
The investigation of optical conductivity in SD materials should lead to some new interesting physics by virtues of their unique low energy spectrum. To understand the direct inter-band optical conductivity, we plot the different constant Fermi energy contours in Fig. [\ref{Fig1}($b$)], with the condition $\delta_0>0$. 
The vertical green arrows/lines in Fig. [\ref{Fig1}($a$ \& $b$)] depict the possible particle-hole direct transitions 
with conserved momentum vector. One can easily identify the avalanches of ${\bf k}$-states available for particle-hole transitions in between particle states with energy $E_+$ and hole states with energy $E_-$, [$E_\pm=E_\pm({\bf k}=0)$] (part ($iii$) of Fig. [\ref{Fig1} ($b$)]). This will result in an abruptly large inter-band joint density of states (JDOS) when the frequency of light corresponds to the difference in energy $\Delta_{{\bf k}=0}=E_+-E_-$ (Fig. [\ref{Fig1} ($c$)]). $\Delta_{{\bf k}=0}$ entirely depends on the gap parameter. 
\\
\indent
The abruptly large inter-band JDOS explained above dictates the giant inter-band optical conductivity at $\hbar\omega=\Delta_{{\bf k}=0}$.
Such giant optical conductivity was also discovered in three-dimensional topological Dirac \textit{semi}-metals\cite{FermiArc,thin}, where the electron-hole transition across the \textit{Fermi arc} contours lead to the very large optical response. 
We investigate the optical conductivity of SD materials by
considering the $x$- and $y$-polarized light, separately. The main finding of this paper is centered around the giant inter-band optical conductivity which interestingly is present only along the $y$-direction while being significantly suppressed along the $x$-direction. 
Immediately, this suggests that such materials should show a relatively high degree of direction-dependent optical transparency. 
It is even more interesting that in one particular direction, at a particular light frequency, the optical transmission is almost blocked. 
\\
\indent
We consider a SD materials subjected to zero-momentum 
electric field ${\bf E} \sim \hat {\bs \nu} E_0 e^{i \omega t} $ ($ \hat {\bs \nu} = \hat {\bf x}, \hat {\bf y} $) with
oscillation frequency $\omega$.
The total charge optical conductivity tensor is given by the relation
$\Sigma_{\nu \xi}(\omega) = \delta_{\nu\xi} \sigma_{D}(\omega) + 
\sigma_{\nu\xi}(\omega)$,
$\sigma_{D}(\omega) = \sigma_d/(1- i \omega \tau)$ is 
the dynamic Drude conductivity due to the intra-band transitions, with 
$\sigma_d$ being the static Drude conductivity and 
$\sigma_{\nu \xi}(\omega) $ is the complex optical conductivity due to
inter-band transitions between particle and hole states.
The real part of the complex optical
conductivity is directly tied to the absorption of the incident
photon energy. It is one of the important tools for
extracting the shape and nature of the material's band dispersion.
The optical conductivity has been extensively studied in various 2D-Dirac materials, from graphene\cite{graphene-op-con,graphene-op-con1,op-con-exp}, silicene \cite{silicene-op-con,silicene-op-con1,silicene-op-con2}, MoS$_2$ \cite{mos2-op-con,mos2-op-con1}, WSe$_2$\cite{WSe2,TAhir}, 8-$Pmmn$ borophene \cite{borophene-op} to the surface
states of topological insulators \cite{ti-op-con,ti-op-con1,ti-op-con2}.
\\
\indent
{\bf Inter-band optical conductivity}:   
Within the framework of linear response theory, 
the Kubo formula for the optical conductivity tensor $\sigma_{\nu\xi}(\omega)$ 
is given by
\begin{eqnarray} \label{kuboG}
{}& \sigma_{\nu\xi}(\omega)=i\frac{e^2}{\omega}\frac{1}{(2\pi)^2}
\int {\bf dk} \nonumber\\& T\sum_n \textrm{Tr} \langle \hat{v}_\nu \hat{G}({\bf k},\omega_n) 
\hat{v}_\xi \hat{G}({\bf k},\omega_n+\omega_l) \rangle_{i\omega_n 
\rightarrow \omega + i\delta}.
\end{eqnarray}
Here $T$ is the temperature and $\omega_l=(2l+1) \pi T$ 
and $\omega_n = 2n \pi T$ are the fermionic and bosonic Matsubara frequencies,  respectively, with $n$ and $l$ being integers.
\begin{figure}[b!] 
	\begin{center}\leavevmode
		\includegraphics[width=70mm,height=55mm]{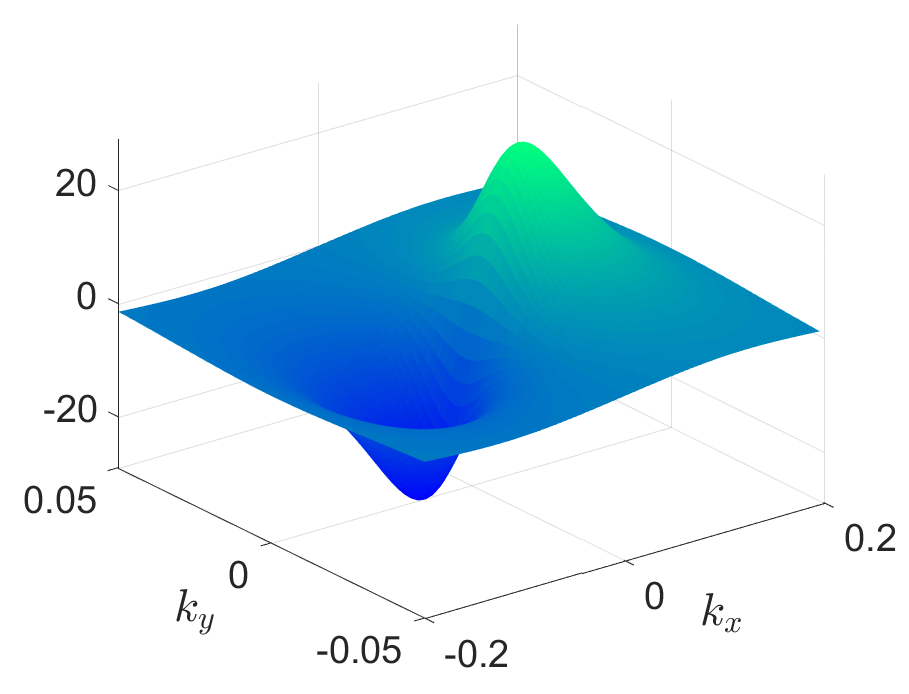}
		\setlength{\belowcaptionskip}{-14pt}
		\caption{Asymmetric Berry curvature of SD Hamiltonian with the perturbation $\delta H=m_0 \sigma_z$.}
		\label{Fig2C}
	\end{center}
\end{figure}

In general, we will consider the effect of perturbation that opens up a gap at the two nodal points. Thus the total hamiltonian is $
H=H_0+\delta H$, where $\delta H=m_0\sigma_z$. The Chern number for this effective Hamiltonian is
$C=\int d {\bf k}\; \Omega({\bf k})$,
where the Berry curvature $\Omega({\bf k})$, \begin{eqnarray}
\Omega({\bf k})=\frac{\bf g}{2|{\bf g}|^3}\cdot \Big(\frac{\partial {\bf g}}{\partial k_x}\times \frac{\partial {\bf g}}{\partial k_y}\Big)=\frac{2\alpha v m_0 k_x}{(g_x^2+g_y^2+m_0^2)^{3/2}},
\end{eqnarray} being asymmetric since $\Omega(k_x,k_y)=-\Omega(-k_x,-k_y)$ (Fig. [\ref{Fig2C}]) would leads to the Chern number being zero. Thus the SD states described by Eq. (\ref{Hamil}) have a trivial topology with Chern number $C=0$\cite{kush}, hence one can predict the transverse conductivity to be zero.
\\
\indent
The corresponding equilibrium Green's function for the modeled Hamiltonian of SD material in Eq. (\ref{Hamil}) is
\begin{eqnarray}\label{green}
\hat{G}({\bf k},\omega) = \frac{1}{2}\sum_\lambda \Big[ \sigma_0 + \lambda\frac{\boldsymbol{\mathcal{F}}\cdot\boldsymbol{\sigma}}{\sqrt{m_0^2+g_y^2+g_x^2}}\Big]G_\lambda({\bf k},\omega),
\end{eqnarray}
where $\sigma_0$ is the unit $2\times 2$ matrix, $\boldsymbol{\sigma}\equiv (\sigma_x,\sigma_y,\sigma_z)$ are the $x$, $y$ and $z$-components of Pauli's matrix and the vector $\boldsymbol{\mathcal{F}}\equiv (\mathcal{F}_x,\mathcal{F}_y,\mathcal{F}_z)$ is given by
\begin{eqnarray}
(\mathcal{F}_x,\mathcal{F}_y,\mathcal{F}_z)=\frac{(\alpha k_x^2-\delta_0,v k_y,m_0)}{\sqrt{(v k_y)^2+m_0^2+(\alpha k_x^2-\delta_0)^2}},
\end{eqnarray}
with 
$ G_\lambda({\bf k},\omega) = [i\hbar\omega + \mu - E_\lambda({\bf k})]^{-1} $.
With the above Green's function in Eq. (\ref{green}), the following quantity, ${\rm Tr}\langle \hat{v}_y\hat{G}({\bf k},\omega_n)\hat{v}_y\hat{G}({\bf k},\omega_n+\omega_l)\rangle$ is obtained as
\begin{eqnarray}
& &\textrm{Tr}\langle\hat{v}_y\hat{G}({\bf k},\omega_n)\hat{v}_y 
\hat{G}({\bf k}, \omega_n + \omega_l)\rangle =2\Big(\frac{v}{\hbar}\Big)^2 \nonumber\\&&\sum_{\lambda,\lambda^\prime}
\Big[1-\lambda \lambda^\prime(\mathcal{F}_x^2-\mathcal{F}_y^2+\mathcal{F}_z^2)
\Big]G_\lambda({\bf k},\omega_n)
G_{\lambda^{\prime}}({\bf k},\omega_l+\omega_n).\nonumber
\end{eqnarray}
Using the Matsubara frequency summation identity
\begin{eqnarray}\label{sumM}
&&T \sum_n \bigg[\frac{1}{i\hbar\omega_n+\mu-E_\lambda} \cdot 
\frac{1}{i\hbar(\omega_l+\omega_n)+\mu-E_{\lambda^\prime} }\bigg]\nonumber\\&&= 
\begin{cases}
\frac{f(E_\lambda)-f(E_{\lambda^\prime})}{i\hbar\omega_l-E_{\lambda^\prime}+E_{\lambda}},& \text{if } 
\lambda\neq \lambda^\prime\\
    0,              & \text{otherwise,}
\end{cases}
\end{eqnarray}
with $f(E)=1/(\exp[\beta(E-\mu)]+1)$ being the fermi Dirac distribution function, $\mu$ is the chemical potential and $\beta=1/k_B T$,
one can write
\begin{eqnarray}\label{trace}
&&T \sum_n \textrm{Tr}\langle \hat v_y \hat{G}({\bf k},\omega_n) \hat v_y 
\hat{G}({\bf k},\omega_l+\omega_n) \rangle =
2\Big(\frac{v}{\hbar}\Big)^2\nonumber\\
&&\sum_{\lambda,\lambda^\prime}\Big[1-\lambda\lambda^\prime(\mathcal{F}x^2-\mathcal{F}_y^2+\mathcal{F}_z^2)\Big]\frac{f_\lambda({\bf k}) - f_{\lambda^\prime}({\bf k})}{i\hbar\omega_l-E_{\lambda^\prime}({\bf k}) + E_\lambda({\bf k})}.\nonumber
\end{eqnarray}
For simplicity, we denoted $f_\lambda({\bf k})\equiv f[E_\lambda({\bf k})]$. 
Using the result of the above equation into Eq. (\ref{kuboG}), we have
\begin{eqnarray} 
\sigma_{yy}(\omega)&=&\frac{e^2}{i 2\pi^2\omega}\Big(\frac{v}{\hbar}\Big)^2\int  {\bf dk}\Big[1-\lambda\lambda^\prime(\mathcal{F}_x^2-\mathcal{F}_y^2+\mathcal{F}_z^2)\Big]
\nonumber\\
& &\frac{f_\lambda({\bf k}) - f_{\lambda^\prime}({\bf k})}
{i\hbar\omega_l-E_{\lambda^\prime}({\bf k})+E_{\lambda}({\bf k})}\Big\vert_{i\omega_l\rightarrow\omega+i\delta}.
\end{eqnarray}
\begin{widetext}
The real part of the optical conductivity which is directly tied to the absorptive part is then given by
\begin{eqnarray}
{\rm Re} \; [\sigma_{yy}(\omega)] =\frac{e^2}{2\pi^2\omega}\frac{v^2}{\hbar^2}\sum_{\lambda,\lambda^\prime}\int {\bf dk}\Big[1-\lambda\lambda^\prime(\mathcal{F}_x^2-\mathcal{F}_y^2+\mathcal{F}_z^2)\Big]
\Big[ f_\lambda({\bf k}) - f_{\lambda^\prime}({\bf k})\Big] \;{\rm Im}\Big[\frac{1}{\hbar \omega+i\delta-E_{\lambda^\prime}({\bf k})+E_{\lambda}({\bf k})}\Big].
\end{eqnarray}
\end{widetext}
Using the identity, ${\rm Im}\Big[\frac{1}{\hbar \omega+i\delta-E_{\lambda^\prime}({\bf k})+E_\lambda({\bf k})}\Big]=-\pi\delta[\hbar\omega-E_{\lambda^\prime}({\bf k})+E_\lambda({\bf k})]$, the above equation is simplified to 
\begin{eqnarray}
&&{\rm Re} \; [\sigma_{yy}(\omega)] =\frac{e^2}{2\pi\omega}\Big(\frac{v}{\hbar}\Big)^2\sum_{\lambda,\lambda^\prime}\int {\bf dk}\Big[ f_\lambda({\bf k}) - f_{\lambda^\prime}({\bf k})\Big]\nonumber\\
&& \Big[1-\lambda\lambda^\prime(\mathcal{F}_x^2-\mathcal{F}_y^2+\mathcal{F}_z^2)\Big]\;\delta[\hbar\omega-E_{\lambda^\prime}({\bf k})+E_\lambda({\bf k})].
\end{eqnarray}
Before we proceed further, Eq. (\ref{sumM}) clearly shows that there can only be inter-band contributions to $\sigma_{\nu\nu}$. The summation over $\lambda$ and $\lambda^\prime$ leaves us with one of the terms that involves $\delta[\hbar\omega-E_-({\bf k})+E_+({\bf k})]$. For evaluating $\rm{Re}\;[\sigma_{yy}(\omega)]$, we don't consider this particular term since it contradicts the conservation of energy. Therefore finally, we have the expression of $\sigma_{yy}$ as follows
\begin{eqnarray}\label{yy}
{\rm Re} \;[\sigma_{yy}(\omega)] =&&\frac{e^2}{2\pi\omega} \Big(\frac{v}{\hbar}\Big)^2\int {\bf dk}\Big[1+\mathcal{F}_x^2-\mathcal{F}_y^2+\mathcal{F}_z^2\Big]\nonumber\\
&&\times\Big[ f_-({\bf k}) - f_{+}({\bf k})\Big] \;\delta(\hbar\omega-\Delta_{\bf k}). 
\end{eqnarray}
Similarly, the real part of $xx$-component of the optical conductivity can be obtained as
\begin{eqnarray}\label{xx}
{\rm Re} \; [\sigma_{xx}(\omega)] =&&\frac{2e^2}{\pi\omega}\Big(\frac{\alpha}{\hbar}\Big)^2\int {\bf dk}(k\cos\theta)^2\Big[1-\mathcal{F}_x^2+\mathcal{F}_y^2+\mathcal{F}_z^2\Big]\nonumber\\
&&\times\Big[ f_-({\bf k}) - f_{+}({\bf k})\Big] \;\delta(\hbar\omega-\Delta_{\bf k}). 
\end{eqnarray}
Here $\theta=\tan^{-1}(k_y/k_x)$ is the azimurthal angle. In our calculation, the effect of impurities is being ignored. It is thus, valid in the ballistic regime, which has been reached in the case of 2D material such as high-mobility suspended or encapsulated graphene\cite{Cala,Bans,wangL}.
\\
\textbf{Drude conductivity}: In order to obtain the total optical conductivity in the entire frequency range, we calculate the static Drude conductivity that dominates in the frequency limit ($\omega\rightarrow 0$). In the framework of Boltzmann-equation, the Drude conductivity is written as 
\begin{eqnarray}\label{drWt}
\sigma_{\nu\nu}^d=-\frac{1}{(2\pi)^2}\int \textbf{dk} |v_{{\bf k}\nu}|^2\frac{\partial f(E({\bf k})}{\partial \;E({\bf k})},
\end{eqnarray}
where $ v_{{\bf k}\nu}=\langle \psi_{\bf k}|\frac{\partial H}{\partial k_\nu}|\psi_{\bf k}\rangle$ is the band velocity along a $\nu$ direction. In the low temperature limit, one can approximate $-\frac{\partial f(E({\bf k})}{\partial \;E({\bf k})}=\delta(E({\bf k})-\mu)$ for evaluating $\sigma_{\nu\nu}^d$.
\\
\begin{figure*}[ht]
	\begin{center}\leavevmode
		\includegraphics[width=130mm,height=58mm]{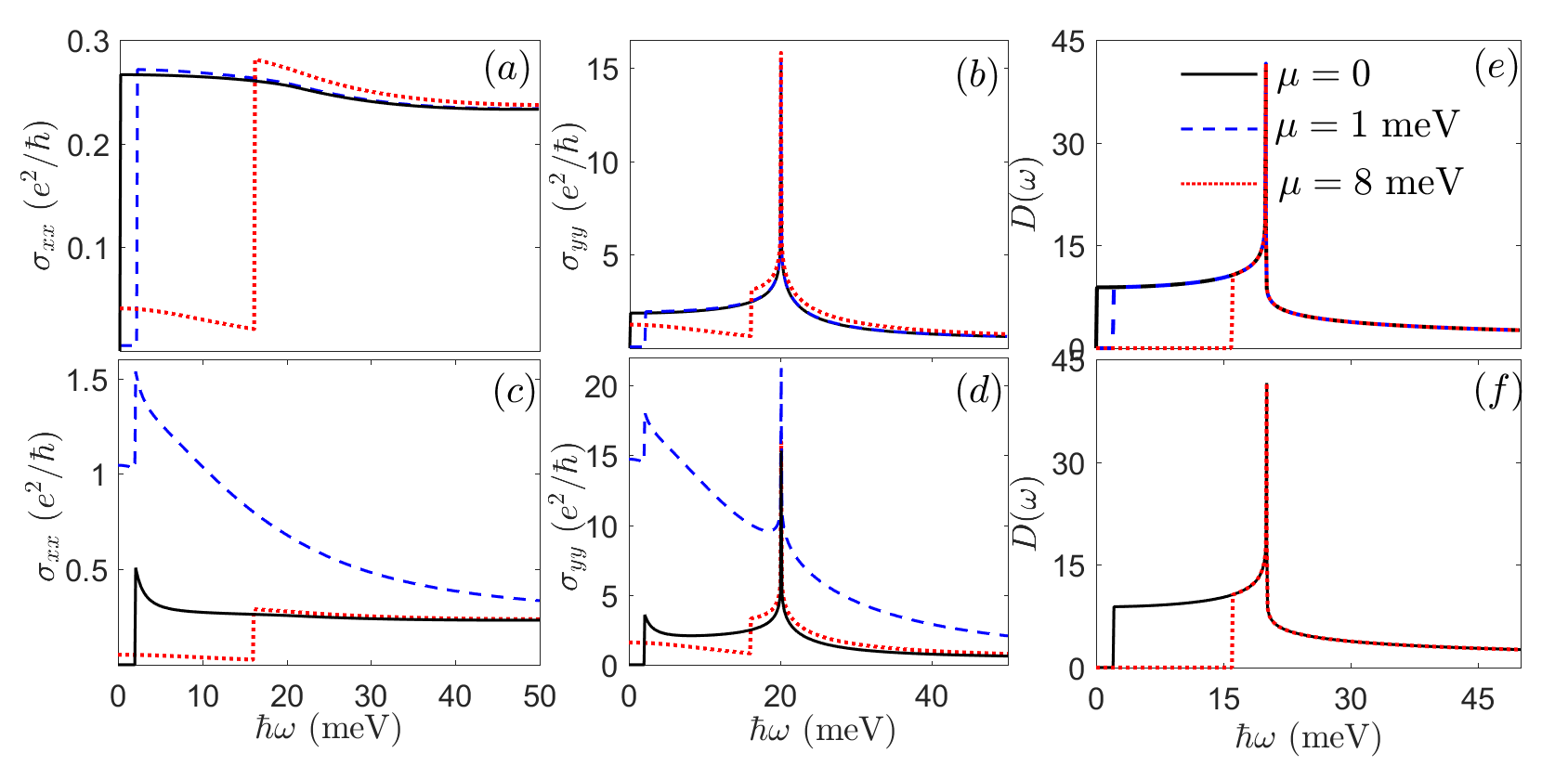} 
		\setlength{\belowcaptionskip}{-12pt}
		\caption{Plots of the real part of $\sigma_{xx}(\omega)$ and $\sigma_{yy}(\omega)$ for gap-less system (Sec. $(a)$ and $(b)$) and gapped system (Sec. $(c)$ and $(d)$) with the corresponding JDOS in Sec. $(e)$ (for gap-less system) and $(f)$ (for gapped system). As shown in legend of Sec. (e), solid black curve, dashed blue and dotted red curve corresponds to $\mu=0$, $\mu=1$ meV and $\mu=8$ meV, respectively for all plots.}
		\label{Fig2}
	\end{center}
\end{figure*}
\indent
For our numerical analysis, we have taken the system parameters for typical SD material according to Ref. [\onlinecite{kush}] and the references therein. The various parameters are taken as $\alpha=7.5$ meV nm$^2$, $v=65$ meV nm, $\delta_0=10$ meV. The SOC induced effective mass that opens up a gap, $m_0=1$ meV and scattering time,
$\tau = 0.04$ ps are taken for illustration purposes.
Firstly, we analyze the inter-band contribution to the optical conductivity given in Eqs. (\ref{yy} \& \ref{xx}). Without the loss of generality, we will discuss the behavior of optical conductivity for the case when the perturbation $\delta H=m_0\sigma_z$ is included. Here, the gap between the valence and conduction bands at the two Dirac points $(k_x,k_y)=(\pm  k_0,0)$ is $2m_0$ and is $\Delta_{{\bf k}=0}=E_+-E_-=2\sqrt{m_0^2+\delta_0^2}$ at ${\bf k}=0$, where $E_\pm=E_\pm({\bf k}=0)=\pm\sqrt{m_0^2+\delta_0^2}$. The limit of $m_0\rightarrow 0$ will yield its behavior for a gap-less SD state. The plots of the total optical conductivity for both gapped and gap-less SD system at three different chemical potential is given in Fig. [\ref{Fig2}]. We consider only lightly doped system, where the chemical potential is always chosen such that $\mu<\Delta_{{\bf k}=0}$. The inter-band optical conductivity spectrum originates at $\hbar\omega=2\mu$. For frequency where $\hbar\omega<2\mu$ and $\hbar\omega<2m_0$, all inter-band transitions are Pauli's blocked (black arrows in Fig. [\ref{Fig1} ($a$)]).  
In the regime $\hbar\omega>2\mu$, there is a smooth variation of its $xx$-component. However, we find that $yy$-component of optical conductivity interestingly, acquires a giant value 
when the frequency of light $\hbar\omega=\Delta_{{\bf k}=0}$. For proper understanding of such features, we again refer to Fig. [\ref{Fig1}($b$ \& $c$)]. In Fig. [\ref{Fig1}($b$)], we show a general representation of the various possible direct inter-band transitions with same ${\bf k}$-vector magnitude across three different constant energy levels. 
The maximum available ${\bf k}$ states occurs when $\hbar\omega=\Delta_{{\bf k}=0}$ (Sec. $(iii)$ of Fig. [\ref{Fig1} ($b$)]). This anomaly results to a giant optical conductivity at $\hbar\omega=\Delta_{{\bf k}=0}$ as seen in Fig. [\ref{Fig2} \& \ref{Fig3}]. The frequency that excites the giant optical conductivity is independent of the chemical potential as shown by the horizontal yellowish line in Fig. [\ref{Fig3} ($(a)$ \& $(b)$)]. The giant optical conductivity $\sigma_{yy}$, implies a huge absoprtion rate for $y$-polarized light of frequency $\hbar\omega=\Delta_{{\bf k}=0}$. Note that, the effect of impurities should lead to indirect inter-band transitions that can slightly suppress the giant optical conductivity.
\\
\indent
A physical insight of the large degree of anisotropy in the optical conductivity also lies in the fact that it has a velocity dependent term ($|\langle +|v_\nu|-\rangle|^2$ that can be simplified from Eq. (\ref{kuboG})) that takes a quasi-particle from the hole to the particle state or vice versa,%
\begin{eqnarray}
\langle -|v_\nu|+\rangle({\bf k})=\begin{cases}-i{2 \alpha v k_x k_y}/{\sqrt{g_x^2+g_y^2}} \;\; {\rm if \;}\nu=x\\
i{ v (\alpha k_x^2-\delta_0)}/{\sqrt{g_x^2+g_y^2}} \;\; {\rm if \;}\nu=y
\end{cases}.
\end{eqnarray}
The above term, with the JDOS is then integrated over the energy contour at $\hbar\omega=\Delta_{\bf k}$.
In the limit of ${\bf k}\rightarrow 0$ (where there is a maximum accumulation of states), the term $|\langle +|v_y|-\rangle|^2=v^2$ whereas the other velocity term vanishes.
\\ 
\indent 
The JDOS (Fig. [\ref{Fig1}($c$), \ref{Fig2}($(e)$ \& $(f)$), \ref{Fig3}($c$)]) which is directly related to the inter-band optical conductivity provides an intuitive way to understand the giant  conductivity at $\hbar\omega=\Delta_{{\bf k}=0}$. The expression of JDOS is
\begin{eqnarray}
D(\omega)=-\frac{1}{4\pi}\sum_{\xi}\int &&d\theta [f_+(k_{\omega,\zeta}(\theta))-f_-(k_{\omega,\zeta}(\theta))]\nonumber\\
&&\times\frac{\delta(k-k_{\omega,\zeta}(\theta))}{|\frac{\partial}{\partial k}(\hbar\omega-\Delta_{\bf k})|_{k_{\omega,\zeta}(\theta)}},
\end{eqnarray}
where $k_{\omega,\zeta}(\theta)$ are the solutions of the equation $\hbar\omega=\Delta_{\bf k}$,
\begin{eqnarray}\label{kw}
k_{\omega,\zeta}(\theta)=\frac{\sec\theta}{\sqrt{2}}\sqrt{\frac{2\delta_0}{\alpha}-\frac{v^2}{\alpha^2}\tan^2\theta\pm \kappa^2},
\end{eqnarray} with $\kappa^2=\sqrt{\frac{(\hbar\omega)^2-4m_0^2}{\alpha^2}-\frac{v^2}{\alpha^2}\Big(4\frac{\delta_0}{\alpha}\tan\theta-\frac{v^2}{\alpha^2}\tan^4\theta\Big)}$ and the subscript $\zeta$ goes for $\pm$ in Eq. (\ref{kw}), which shows that there could be two values of $k_\omega(\theta)$, say $k_{\omega,1}$ for ``$+$" and $k_{\omega,2}$ for ``$-$". It is easy to see from Fig. [\ref{Fig1}($c$)], that we have both solutions $k_{\omega,1}$ and $k_{\omega,2}$ only for $\hbar\omega<\Delta_{{\bf k}=0}$, while there is only one possible $k_{\omega,1}$ value for $\hbar\omega>\Delta_{{\bf k}=0}$.
\\ \indent
In Fig. [\ref{Fig3} $(c)$], separating the regime $I$ and $II$ is the line $\hbar\omega=2\mu$, with $\hbar\omega>=2m_0$. Below this line is region $I$ which includes the parts with all the Pauli's blocked inter-band transition (indicated by black arrows in Fig. [\ref{Fig1} $(a)$]), which thus leads to zero JDOS in this region and ultimately zero inter-band optical conductivity. The JDOS is finite in regime $II$ and $III$. Separating the regime $II$ and $III$ is a yellowish horizontal line with $\hbar\omega=\Delta_{{\bf k}=0}$. This line corresponds to the giant optical conductivity  which is independent of the chemical potential. This anomaly in the JDOS arises from the van Hove singularity which as indicated in the above discussion, is due to the maximum accumulation of states at ${\bf k}=0$.
\begin{figure}[ht]
	\begin{center}\leavevmode
		\includegraphics[width=85mm,height=70mm]{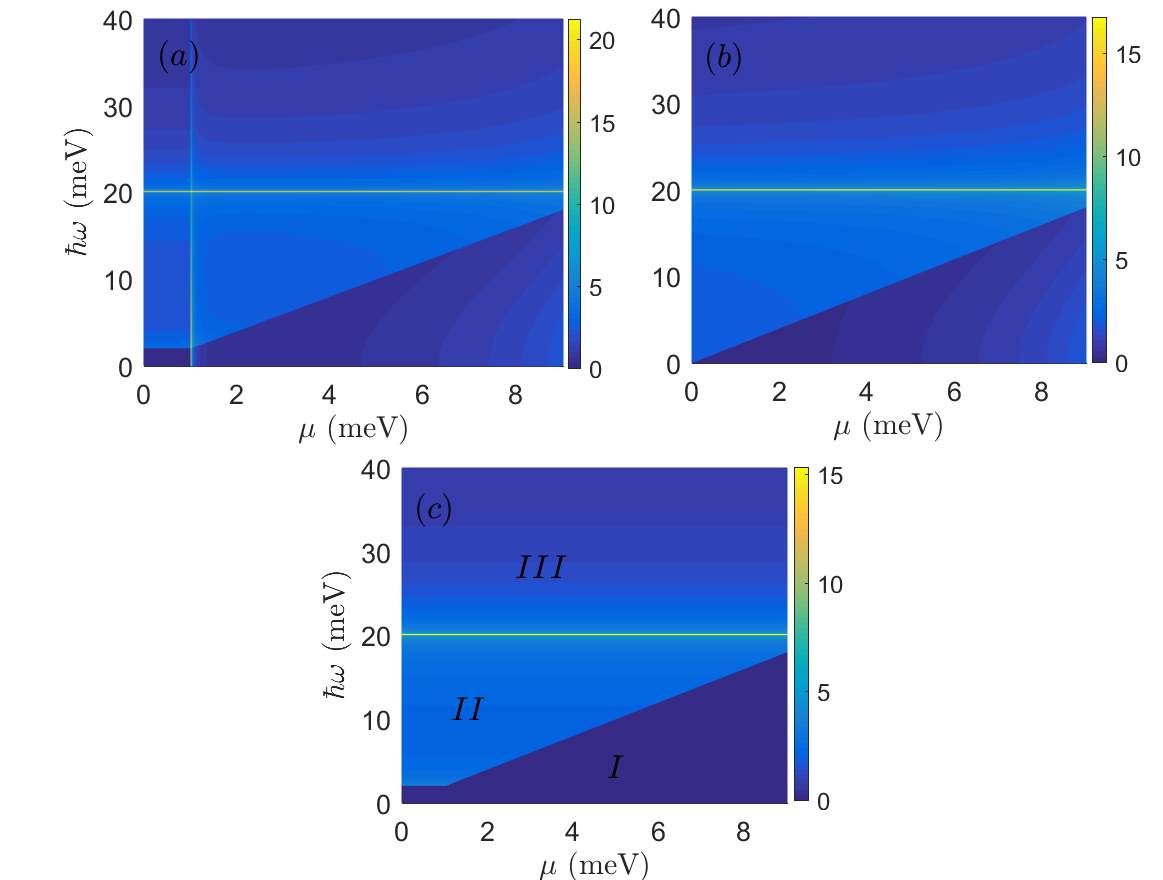}
		\setlength{\belowcaptionskip}{-24pt}
		\caption{Gradient plot of $yy$-components of conductivity (in units of $e^2/\hbar$) for (a) $m_0=1$ meV, (b) $m_0=0$ meV, and (c) JDOS. The giant optical conductivity is only excited by frequency $\hbar\omega=\sqrt{m_0^2+\delta_0^2}$.}
		\label{Fig3}
	\end{center}
\end{figure}
\\ 
\indent 
In \textit{conclusions}, we have presented detailed theoretical studies of the optical conductivity of 2D \textit{semi}-Dirac materials. As an influence of the gap parameter, $\delta_0$, we found that the optical response is highly sensitive to the direction of polarization of light. For light polarized in the direction where the dispersion is linear, our results predict a giant inter-band optical conductivity when the frequency corresponds to the electron-hole states energy separation at ${\bf k}=0 $, while on the other hand, the inter-band optical conductivity is significantly suppressed when light is polarized along the direction orthogonal with respect to the former. Also, the frequency that excites this giant optical conductivity is found to be independent of the chemical potential for lightly doped \textit{semi}-Dirac system. The high degree of anisotropy of optical conductivity suggested that the SD materials can be a potential candidate of a unique transparent conductor with a given transparency along one direction while bearing a very high absorption rate along the orthogonal direction.
Also, the direction dependency of this giant inter-band optical conductivity can be presented as a tool that can be uniquely used to probe the dispersive nature of 2D \textit{semi}-Dirac materials. To wind up this paper, we also proposed the possibility of extracting some interesting physics that may co-exist along with the giant optical conductivity in super-crystal (TiO$_2$)$_5$/(VO$_2$)$_3$ which shows Chern insulating states, where the band dispersion is governed by a ``type-II" \textit{semi}-Dirac dispersion.
\\
\indent
The authors would like to thank SK Firoz Islam for useful discussions throughout this work. This work is an outcome of the Research and Development work undertaken in the project under the Visvesvaraya PhD Scheme of Ministry of Electronics and Information Technology, Government of India, being implemented by Digital India Corporation (formerly Media Lab Asia).

\end{document}